# High-efficiency broad-bandwidth sub-wavelength grating based fibre-chip coupler in SOI

Siddharth Nambiar[a], Shankar K Selvaraja*[a]

[a]Center for Nanoscience and Engineering (CeNSE), Indian Institute of Science (IISc), Bengaluru, India, 560012

Abstract. We report a comprehensive study on the performance of uniform and non-uniform based subwavelength grating couplers in Silicon-on-Insulator. The two performance metrics, coupling efficiency and bandwidth enhancement and trade-offs is presented. We also present, design parameters to achieve high-efficiency and broadband operation based on the detailed study of various loss mechanisms in the grating. Based on the detailed analysis subwavlength grating coupler designs with efficiency as high as 84 % with a 1 dB bandwidth of 50 nm and 98 % with a 44 nm 1 dB bandwidth is presented.

* shankarks@iisc.ac.in

1 Introduction

One of the critical issues in integrated photonic circuits in silicon-on-insulator (SOI) platform is the ability to efficiently couple broadband light between a single-mode optical fiber (SMF) and the *Si* photonic wire waveguide. Edge- or butt-coupling using sophisticated mode converters have been proposed and demonstrated with high-coupling efficiency (<1 dB) and broadband operation [1– 3]. Such solutions are, however, demanding cumbersome processing steps and are not compatible with CMOS integration scheme. Alternatively, out-of-plane coupling using diffraction gratings is an elegant and versatile technique for coupling light between a sub-micron waveguide and a SMF [4, 5].

Even though grating coupler is simple in operation and implementation, the coupling efficiency and bandwidth of operation are limited. The coupling efficiency can be improved by either increasing its directionality to the optical fiber and/or by improving the modal overlap between the fiber and the diffracting grating [6]. On the other hand, improving bandwidth is not trivial [7]. One need to engineer the grating dispersion for bandwidth improvement. While most of the surface grating couplers have been implemented by partial etch in the *Si* device layer, it has limited flexibility in achievable refractive index modulation and also desired yield due to etch depth non-uniformity.

Recently, there have been multiple proposals and demonstrations made to use subwavelength grating (*SWG*) for coupling light[8–10]. A subwavelength structure essentially consists of an artificially engineered material that due to its subwavelength dimensions behave like a homogeneous medium. These structures provide a convenient tool for tuning the average effective index of a grating as compared to the shallowly etched grating. Furthermore, they can be fabricated by fully etching the *Si* device layer using buried oxide in SOI as an etch stop layer makes it tolerant to etch non-uniformity.

In this paper, we present subwavelength grating designs based on binary i.e. *Si-SWG* where the high index region is composed of *Si* and the low-index region of the subwavelength structure, as well as *dual-SWG* gratings in which both high and low index regions are composed of subwavelength structures. We present a comprehensive study on the design trade-off's between coupling efficiency and bandwidth and also show how one could improve upon them by using Bragg mirror at the bottom of the grating to improve directionality as well as apodization to improve mode overlap between emitted grating field and the fiber field.

2 Numerical analysis

According to the zeroth-order approximation of the effective medium theory, the refractive index of the composit medium consisting of two subwavlength grating of two materials. For a grating with *Si* and *SiO₂* as high and low index medium respectively, the effective index of the TE and

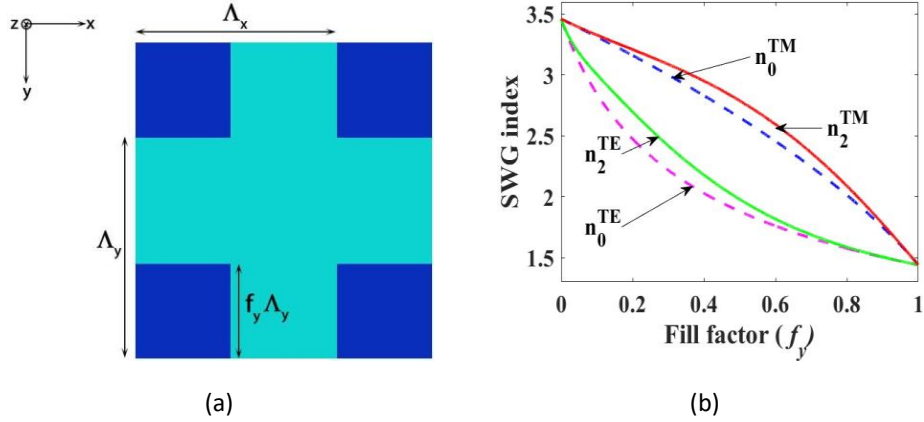

(a) (b)

Fig 1: (a) A typical subwavelength grating coupler with longtitudinal period $\Lambda_x$ and lateral subwavelength period $\Lambda_y$. The dark blue region $f_y\Lambda_y$ signifies the slit width to be filled with *SiO₂*. (b) The variation of subwavelength indices for *TE* and *TM* polarizations with lateral subwavelength fill factor $f_y$ for $\Lambda_y = 300nm$

TM polarised fundamental mode is calculated as [11],

$$\frac{1}{n_0^{TE}} = \left(\frac{f_y}{n_{ox}^2} + \frac{(1-f_y)}{n_{Si}^2}\right)^{\frac{1}{2}} \tag{1}$$

$$n_0^{TM} = \left(f_y n_{ox}^2 + (1-f_y) n_{Si}^2\right)^{\frac{1}{2}} \tag{2}$$

Here $f_y$ is the subwavelength fill factor in the lateral *y* direction as depicted in Fig. 1(a). When the ratio $n_{eff}\Lambda_y/\lambda_0$ approaches 1, higher order approximation should be considered. Accordingly the effect medium refractive index has to be calculated as,

$$n_2^{TE} = n_0^{TE}\left(1 + \frac{\pi^2}{3}\left(\frac{\Lambda_y}{\lambda}\right)^2 f_y^2(1-f_y^2)(n_{ox}^2 - n_{Si}^2)\left(\frac{n_0^{TM}}{n_{eff}}\right)^2\left(\frac{n_0^{TE}}{n_c n_{Si}}\right)^4\right)^{1/2} \tag{3}$$

The average effective grating refractive index is then determined by,

$$n_{effg} = f_x n_{eff_l} + (1 - f_x) n_{eff_h} \tag{4}$$

where $n_l^{eff}$ and $n_h^{eff}$ being the low and high effective TE indices for the subwavelength regions $n_L$ and $n_H$ respectively. As discussed in the earlier section, in addition to coupling efficiency, bandwidth of the coupler is an important performance metric. The intrinsic 1dB bandwidth of grating is given by [12],

$$\Delta \lambda_{1dB} = \eta_{1dB} \left| \frac{-n_c \cos\theta}{\frac{n_g^{eff} - n_c \sin\theta}{\lambda_0} - \frac{d n_{eff}(\lambda)}{d\lambda}} \right| \tag{5}$$

Where $\eta_{1dB}$ is a fiber related coefficient, $\theta$ the incident angle and $\frac{d n_{eff}(\lambda)}{d(\lambda)}$ is the waveguide dispersion. The bandwidth limitation essentially arises due to the fact that change in the wavelength leads to a change in the radiated angle. A fixed inclined fiber is no longer efficiently able to capture the diffracted power reulting in reduced coupling efficiency at those wavelengths. A suitable solution to increase bandwidth would be to decrease the effective grating index $n^{eff}_g$ (5). In a traditional shallowly etched grating, $n_g^{eff}$ can be decreased by increasing the etch depth, however, this will result in large reflection from the grating. An alternative approach is to replace the low refractive index region by a subwavelength grating [7] [13] or implementing a *dual-SWG* altogether. Another factor is that the waveguide dispersion is typically negative in the normal regime which may lead to a reduction in operating bandwidth. One way to address this is by using a lower subwavelength feature [7].

*2.1 Uniform subwavelength grating coupler (SWGC)*

In this section, we present uniform binary *Si-SWG* and *dual-SWG* grating configurations. A schematic of the design layout is shown in Fig. 2. The SOI stack consists of a 220 nm *Si* on top of 2.2 $\mu$m of $SiO_2$ Buried Oxide (BOX). Unlike stadard 2 $\mu$m of BOX we have chosen an optical thickness 2.2 $\mu$m to present the untimate achievable efficiecy and bandwidth. It has been already shown that 2 $\mu$m BOX SOI not ideal for fully etched gratings [14]. The grating is covered with $SiO_2$ as top cladding material. The grating is implemented by periodic interlace of low and high refractive index regions. The grating periodicity $\Lambda_x$ is determined from the phase matching condition as,

$$p\lambda_0 = \Lambda_x(n_g^{eff} - n_c \sin\theta) \qquad (6)$$

where $p$ is the diffraction order (approximated to 1 in this case), $n_c$ is the cladding index, $n^{eff}_g$ is the average effective grating index and $\lambda_0$ is the desired peak wavelength of the grating.

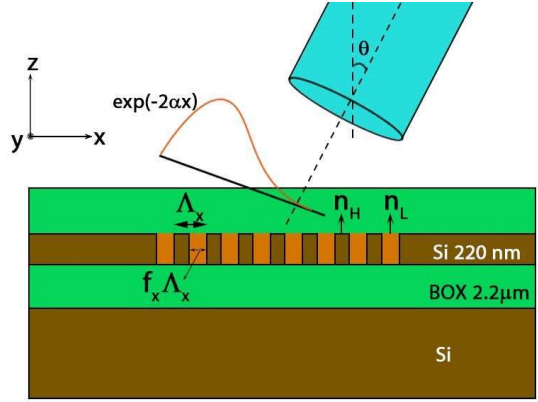

Fig 2: Schematic of a uniform 2D subwavelength grating coupler design used in FDTD simulations. The grating coupler has a fixed period and duty cycle due to which power radiated upwards is exponential.

The characteristics of the grating coupler is simulated using 2D FDTD method. The grating is excited by a TE polarized Gaussian approximated fiber source embedded in the top cladding with an $1/e^2$ intensity beam diameter of 10.4 $\mu$m. For both *Si-SWG* and *dual-SWG* the duty cycle is fixed at 50% with 24 periods. The period of the diffraction grating is calculated for each subwavelength index using Eq.6. The coupling efficiency is determined by placing a power monitor in the waveguide at a sufficient distance from the grating. The lateral position of the source is optimized for each SWG index to achieve maximum coupling. Apart from the coupling efficiency and bandwidth, other critical characteristics such as reflection, transmission, substrate leakage and extraction efficiency are also calculated by exiting the fundamental TE mode of the waveguide.

3 Results and Discussion

Figure 3 summarises the grating performance as a function of lower subwavelength index ratio $n_L$ for *Si-SWG* and *dual-SWG* grating couplers. Typically the power diffracted by the grating can be quantified by four parameters,

$$P = P_{clad} + P_{sub} + P_{refl} + P_{tr} \tag{7}$$

$$P_{CE} = \eta P_{clad} \tag{8}$$

where $P_{clad}$ denotes the power extracted to the top cladding, $P_{sub}$ denotes the substrate leakage, $P_{refl}$ denotes the back reflection and $P_{tr}$ signifies the power transmitted through the grating. $\eta$ is the mode field overlap between the diffracted grating field and the Gaussian fiber mode [4] and is expressed as $\int_S |E_{gr} \times H^*_{fib}|^2$ with *gr* denoting the emitted grating field and *fib*, the fiber mode field. The power coupling efficiency $P_{CE}$ is then a product of $\eta$ and $P_{clad}$. With the examination of the loss mechanisms in the grating, one can find suitable ways for improving the performance.

Figure 3(a) and 3(c) shows the various power loss mentioned above along with the $P_{CE}$ for *SiSWG* and *dual-SWG* couplers respectively. $P_{CE}$ in both cases follow the pattern of extraction to top cladding ($P_{clad}$). $P_{CE}$ for *Si-SWG* is observed to be slightly higher at 56.8% which amounts to a loss of -2.45 dB for a lower subwavelength index $n_L$ of 2.6. An index of 2.6 corresponds to an etch depth of 70 nm on a 220 nm SOI, which agrees with the standard shallow etch implementation of the grating coupler. A lower $n_L$ would mean a lower $n_{eff}$ and hence greater contrast which observed in the case of *Si-SWG* couplers. A higher $n_L$, on the other hand, leads to a weak grating and results in a higher $P_{tr}$. For *dualSWG*, $P_{CE}$ is observed to be 52.8 % amounting to a loss of -2.77 dB. In both the cases, the optimum operating point is a result of optimum field overlap $\eta$ as well as minimal power loss due to back reflection, grating transmission, and substrate leakage. Although the back reflection to the waveguide is <0.3 % at peak CE, the substrate leakage fraction $P_{sub}$ is still high at over 30% for both *Si-SWG* and *dual-SWG*. It is clear from this analysis that most of the power loss is emanating from substrate leakage. As has been observed in previous works, this loss can be mitigated by inserting a bottom reflector at an optimal distance underneath the grating. [13, 15, 16]. Figure 3(b) and 3(d) depicts the evolution of bandwidth for different coupler configuration and SWG indices. A general trend observed is that, for any coupler configuration the bandwidth monotonically increases with decreasing $n_L$ index, which is a result of lower $n_g^{eff}$.

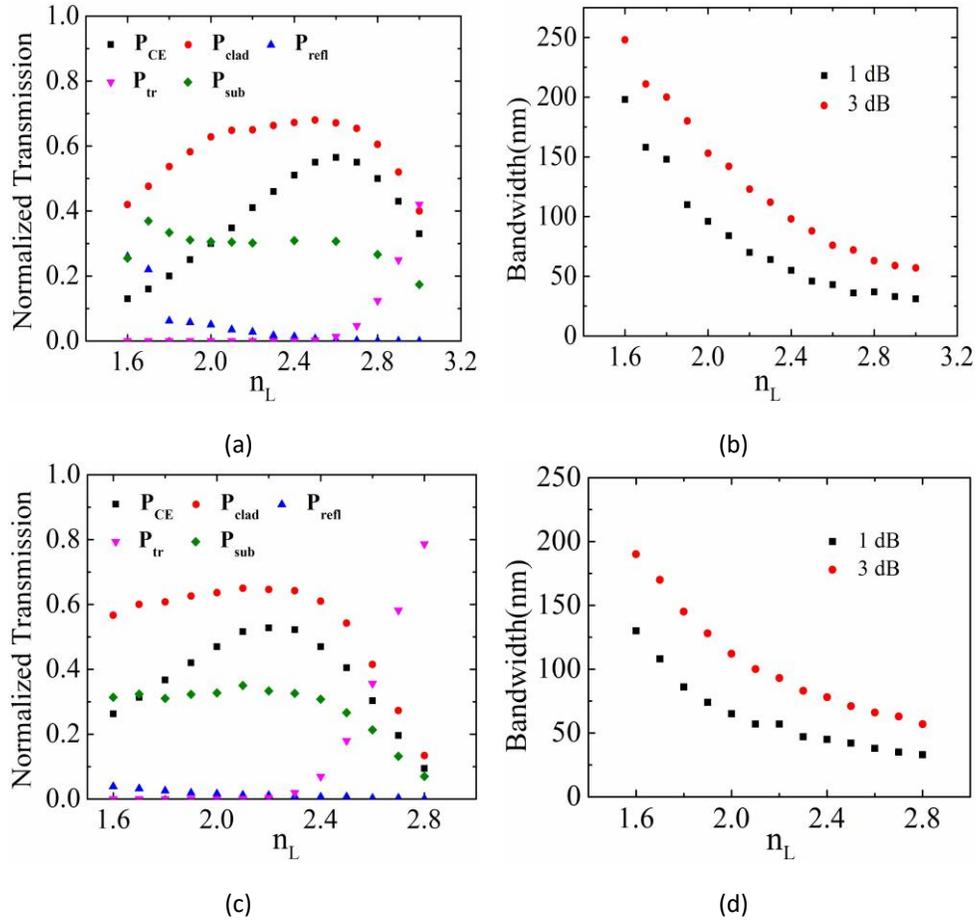

Fig 3: (a), Fiber to waveguide peak coupling efficiency ($P_{CE}$), power extraction to top cladding ($P_{clad}$), substrate leakage ($P_{sub}$), grating back reflection ($P_{refl}$), and transmission through grating ($P_{tr}$) as function of low SWG indices $n_L$ for $n_H$=3.46. All parameters are plotted at wavelength of peak CE. (b) corresponding 1 and 3 dB bandwidth variation as function of $n_L$. (c), (d) the same for $n_H$=3.0. leads to a high back reflection $P_{refl}$, which is more so

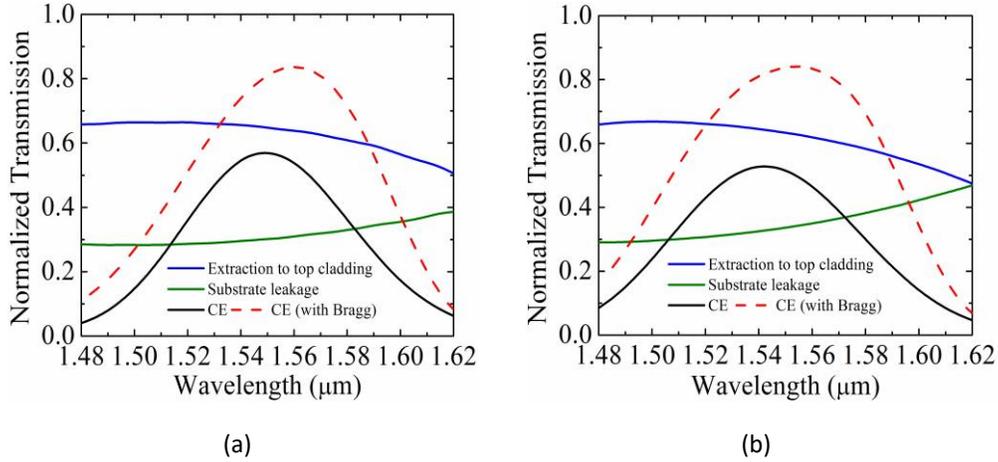

(a)          (b)

Fig 4: Grating spectrum consisting of power extraction to top cladding, substrate leakage as well as coupling efficiency with and without a Bragg mirror for (a) $n_H$=3.46, $n_L$=2.6 and (b) $n_H$=3, $n_L$=2.3

| $n_H$ | $n_L^{peak}$ | CE(-dB) | | 1dB(nm) | | BOX($\mu m$) | | $\Lambda_y$ = 300(nm) | |
|---|---|---|---|---|---|---|---|---|---|
| | | Si Sub | Bragg mirror | Si Sub | Bragg mirror | Si Sub | Bragg mirror | $f_y\Lambda_y$ ($n_H$) | $f_y\Lambda_y$ ($n_L$) |
| 3.46 | 2.6 | 2.45 | 0.77 | 43 | 50 | 2.2 | 2.14 | - | 70 |

| 3   | 2.3 | 2.77 | 0.75 | 48 | 60 | 2.2 | 2.17 | 30 | 105 |
| 2.8 | 2.1 | 3    | 0.88 | 54 | 68 | 2.2 | 2.19 | 50 | 130 |
| 2.6 | 1.9 | 3.26 | 1.16 | 59 | 73 | 2.2 | 2.2  | 70 | 165 |

Table 1: Performance summary of uniform *Si-SWG* and *dual-SWG* grating coupler configurations. CE(in -dB) and bandwidth with Si substrate and with Bragg mirror. Also given are the lithographic features required to implement the corresponding subwavelength indices.

Figure 4 shows the spectral characteristics of the different subwavelength coupler configurations along with the coupling efficiencies with and without a bottom reflector at optimum $n_L$. A 4 layer *Si/SiO₂* Bragg bottom mirror of thickness of 110 nm and 270 nm respectively is considered. The power extraction to top cladding and substrate leakage remains almost identical for both *Si-SWG* and *dual-SWG*. The addition of bottom reflector is seen to substantially improve CE to over 80 % due to constructive interference between top extracted power and substrate reflected power. Table 1 highlights the relative performances of the different grating configurations with a bottom reflector. While the efficiency of *Si-SWG* coupler is found to increase to -0.77 dB, peak CE among all designs with bottom reflector is observed for $n_H$=3; CE of -0.75 dB with a 1 dB bandwidth of 60 nm. In all cases, 1 dB bandwidth is observed to be slightly higher for bottom reflector. This higher bandwidth may be attributed to enhanced off-peak reflection by the mirror leading to a broadening of the passband [17]. A higher 1 dB bandwidth of 73 nm is achieved with $n_H$=2.6, however with a peak CE od 1.16 dB that clearly indicates the CE-bandwidth trade-off.

The physical dimensions of the subwavelength grating to achieve the desired index are shown in Table 1. It has to be noted that the physical dimensions of the grating are well within the capability of advanced optical lithograhy and electron beam lithography systems.

*3.1 Non-Uniform SWGCs*

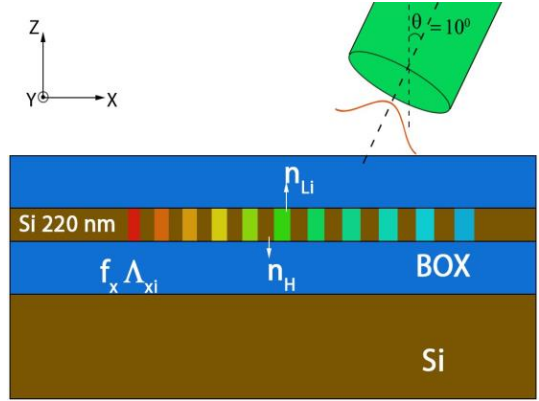

Fig 5: Schematic of non-uniform 2D subwavelength grating coupler design used in FDTD simulations.

As mentioned previously, one of the two primary loss mechanisms is the grating and fiber mode mismatch. For a uniform grating, the power emission profile is an exponentially decaying function ($P_{out} = P_{in}e^{-2\alpha x}$) due to the leakage parameter $\alpha$ being constant in propagation direction. However, a fiber mode has a Gaussian profile, which results in an overlap mismatch with the grating field. The overlap is increased by designing the grating to scatter a Gaussian profile, which requires progressive change in the scattering strength of the grating. For the subwavelength structures, this can be achieved by tuning the period, which in turn modifies the subwavelength index and hence the effective grating index.

A schematic of variable grating used for 2D FDTD is shown in fig.5. The scattered field for a non-uniform grating with linear apodizations is determined by the non-uniform coupling coefficient [18] as,

$$\alpha(x) = \frac{G^2(x)}{1 - \int_0^x G^2(t)dt} \tag{9}$$

where $G(x)$ is the normalized Gaussian profile determined by the fiber mode source and $x$ the propagation direction. The apodization implemented in the case of a *Si-SWG* was a linear chirp function with $n_L$ varied from 1.9 to 3.2 and $n_H$ fixed at 3.46. With a fixed 50% duty cycle, period was also continuously varied. The apodized grating was constructed by modifying eqn.4 as,

$$\sum_{i=1}^{p} n_{gi}^{eff} = \sum_{n_{Li}=n_{L}^{min}}^{n_{L}^{max}=3.2} (fn_{Li}^{eff} + (1-f)n_{H}^{eff})$$

Here $n_L^{eff}$ and $n_H^{eff}$ are the effective indices (TE) of the low and high index subwavelength regions respectively. From the above expression, the longitudinal grating period of the $i^{th}$ element $\Lambda_i$ can be extracted from eqn.6. For the *dual-SWG*, a two step apodization was employed whereby the chirp function is linear upto $n_H$ after which both $n_L$ and $n_H$ are gradually increased till $n_H$ reaches a maximum of 3.2 and $n_L$=3.15. Such a dual stage apodization reduces the back reflection from the subwavelength index region. The grating thus was constructed as,

$$\sum_{i=1}^{p} n_{gi}^{eff} = \sum_{n_{Li}=n_{L}^{min}}^{n_{H}^{min}-0.05} (fn_{Li}^{eff} + (1-f)n_{H}^{min\,eff})$$

$$\sum_{j=p+1}^{q} n_{gj}^{eff} = \sum_{n_{Lj}=n_{H}^{min}+0.05, n_{Hj}=n_{H}^{min}+0.1}^{n_{Lj}=3.15, n_{Hj}=3.2} (fn_{Lj}^{eff} + (1-f)n_{Hj}^{eff})$$

The grating spectrum for these non-uniform couplers is plotted in fig.6. A top cladding extraction of 70 % and substrate leakage of 30 % are similar to that observed in the case of uniform couplers. However, the observed peak CE is observed to be 67 % and 61 % for *Si-SWG* and *dual-SWG* respectively, which is about 10 % higher than the corresponding uniform designs. With a Bragg mirror, the CE for *Si-SWG* is enhanced to 98.25 (-0.077 dB) % and 96.8 (-0.14 dB) % for a *dual-SWG* with $n_H$=3.

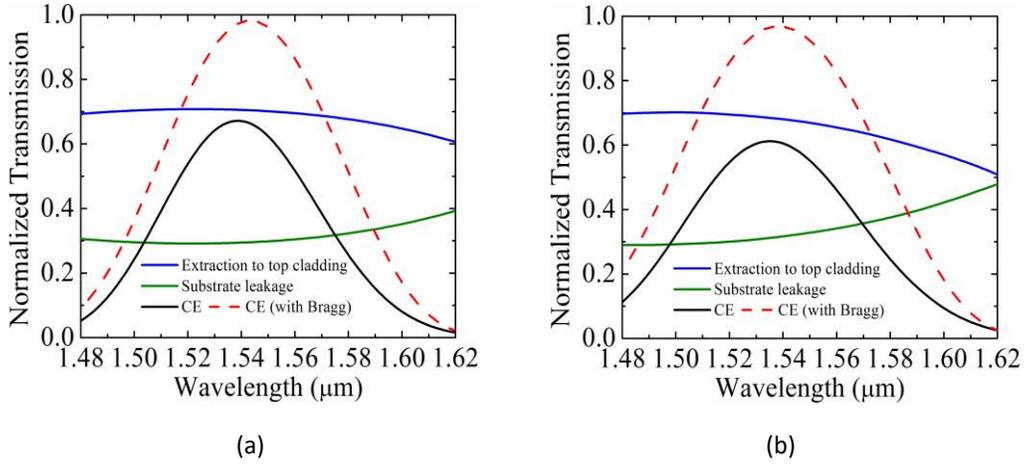

Fig 6: Grating spectrum consisting of power extraction to top cladding, substrate leakage as well as coupling efficiency with and without a Bragg mirror for (a) apodized *Si-SWG* and (b) apodized *dual-SWG* with $n_H$ =3.

Fig.7 shows the normalized E-field intensities for both uniform and apodized couplers. The higher coupling in case of apodized couplers can be attributed to the complete enclosure of the in-cident fiber mode by the grating field which leads to a better mode overlap with the fiber field compared to that of uniform couplers. Table 2 provides a performance summary of different apodized subwavelength grating designs. In all designs, the peak CE is observed to be 10-14 % higher than the corresponding optimized uniform coupler designs. For the design with the lowest CE (i.e $n_H$ = 2.6), the observed 3 dB bandwidth was 95 nm, which is higher than previously reported non-uniform subwavelength coupler designs. One of the crucial parameters in designing SWG is the minimum dimension of the structures. For a desired refractive index, the choice of SWG period has a direct implication on the minimum trench width of the grating. As discussed earlier, large bandwidth requires lower dispersion (Eq. 5) that can achieve by using a lower SWG period. For an SWG period of 300 nm, a minimum trench width of 13.5 nm would be required for the apodized coupler, which could be achieved using e-beam lithography or 193 immersion lithography. The minimum trench width could be increased to 26 nm with 400 nm period. Availability of patterning technology and resolution could limit the SWG period choice.

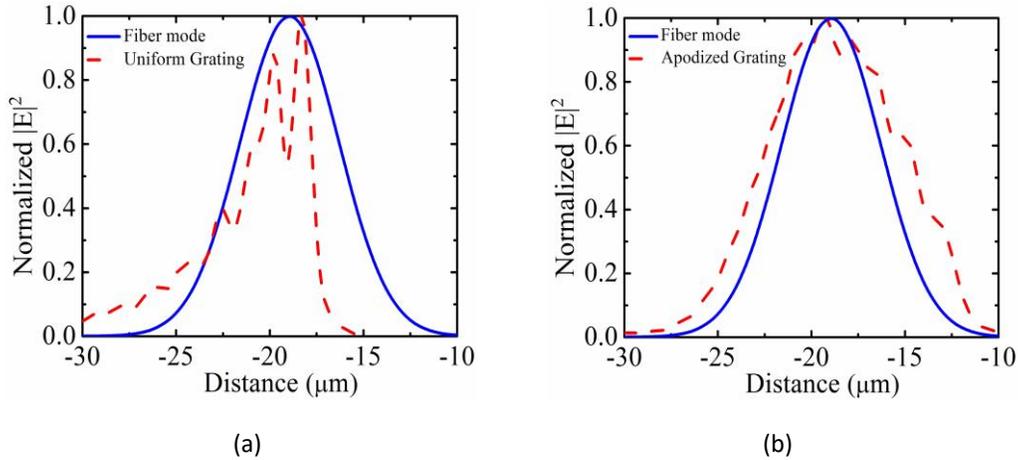

Fig 7: Normalized E field intensities along with the fiber mode field for (a) uniform gratings, (b) apodized gratings.

| $n_H^{max}$ | $n_H^{min}$ | $n_L^{min}$ | CE(-dB) | | 1dB(nm) | | BOX($\mu$m) | |
|---|---|---|---|---|---|---|---|---|
| | | | Si Sub | Bottom mirror | Si Sub | Bottom mirror | Si Sub | Bottom mirror |
| 3.46 | 3.2 | 1.9 | 1.73 | 0.077 | 40 | 44 | 2.2 | 2.15 |
| 3.2 | 3.0 | 1.7 | 2.13 | 0.14 | 44 | 50 | 2.2 | 2.16 |
| 3.2 | 2.8 | 1.65 | 2.48 | 0.241 | 47 | 53 | 2.2 | 2.17 |
| 3.2 | 2.6 | 1.6 | 2.66 | 0.376 | 51 | 57 | 2.2 | 2.18 |
| Ref[19] | - | - | 1.97 | 0.42 | - | 66* | - | - |
| Ref[20] | - | - | - | 0.43 | - | 76* | - | - |

Table 2: Performance summary of apodized *Si-SWG* and *dual-SWG* grating coupler configurations. Max CE (in -dB) and bandwidth with Si substrate and the optimized BOX thickness for a 4 layer bottom Bragg mirror. Comparison with references for simulated data only.* 3dB bandwidth.

4 Conclusion

We have done a detailed studied of a subwavelength grating based couplers with the main objective to explore ways to improve coupling efficiency and operation bandwidth. We observe a trade-off between maximum achievable coupling efficiency and bandwidth. Based on the study, we found that for uniform grating designs, the maximum CE can vary from -2.45 dB to -3.26 dB per coupler with the operating 1 dB bandwidth being 43 nm and 59 nm respectfully. For non-uniform designs the maximum CE can very from -1.73 dB to -2.66 dB with the corresponding 1 dB bandwidth varying from 40 to 51 nm. For uniform designs with a bottom reflector best CE is observed for a *dual-SWG* design at -0.75 dB with a 1 dB bandwidth of 60 nm. For a non-uniform couplers, best CE is observed for *Si-SWG* at -0.077 dB with the 1 dB bandwidth being 44 nm.

References

[1] B. B. Bakir, A. V. de Gyves, R. Orobtchouk, *et al.*, "Low-Loss (¡1 dB) and PolarizationInsensitive Edge Fiber Couplers Fabricated on 200-mm Silicon-on-Insulator Wafers," *IEEE*

*Photonics Technol. Lett.* 22, 739–741 (2010).

[2] M. Pu, L. Liu, H. Ou, *et al.*, "Ultra-low-loss inverted taper coupler for silicon-on-insulator ridge waveguide," *Opt. Commun.* 283(19), 3678–3682 (2010).

[3] V. R. Almeida, R. R. Panepucci, and M. Lipson, "Nanotaper for compact mode conversion," *Opt. Lett.* 28, 1302–1304 (2003).

[4] D. Taillaert, P. Bienstman, and R. Baets, "Compact efficient broadband grating coupler for silicon-on-insulator waveguides," *Opt. Lett.* 29, 2749–2751 (2004).

[5] A. Mekis, S. Gloeckner, G. Masini, *et al.*, "A Grating-Coupler-Enabled CMOS Photonics Platform," *IEEE J. Sel. Top. Quantum Electron.* 17, 597–608 (2011).

[6] G. Roelkens, D. V. Thourhout, and R. Baets, "High efficiency Silicon-on-Insulator grating coupler based on a poly-Silicon overlay," *Opt. Express* 14, 11622–11630 (2006).

[7] X. Xu, H. Subbaraman, J. Covey, *et al.*, "Colorless grating couplers realized by interleaving dispersion engineered subwavelength structures," *Opt. Lett.* 38, 3588–3591 (2013).

[8] X. Chen and H. K. Tsang, "Polarization-independent grating couplers for silicon-on-insulator nanophotonic waveguides," *Opt. Lett.* 36, 796–798 (2011).

[9] R. Halir, A. Ortega-Monux, J. H. Schmid,˜ *et al.*, "Recent Advances in Silicon Waveguide Devices Using Sub-Wavelength Gratings," *IEEE J. Sel. Top. Quantum Electron.* 20, 279–291 (2014).

[10] Q. Zhong, V. Veerasubramanian, Y. Wang, *et al.*, "Focusing-curved subwavelength grating couplers for ultra-broadband silicon photonics optical interfaces," *Opt. Express* 22, 18224–18231 (2014).

[11] D. H. Raguin and G. M. Morris, "Antireflection structured surfaces for the infrared spectral region," *Appl. Opt.* 32, 1154–1167 (1993).

[12] X. Chen, K. Xu, Z. Cheng, *et al.*, "Wideband subwavelength gratings for coupling between silicon-on-insulator waveguides and optical fibers," *Opt. Lett.* 37, 3483–3485 (2012).

[13] W. S. Zaoui, M. F. Rosa, W. Vogel, *et al.*, "Cost-effective CMOS-compatible grating couplers with backside metal mirror and 69% coupling efficiency," *Opt. Express* 20, B238—-B243 (2012).

[14] M. Antelius, K. B. Gylfason, and H. Sohlstrom, "An apodized SOI waveguide-to-fiber surface¨ grating coupler for single lithography silicon photonics," *Opt. Express* 19, 3592–3598 (2011).

[15] S. K. Selvaraja, D. Vermeulen, M. Schaekers, *et al.*, "Highly efficient grating coupler between optical fiber and silicon photonic circuit," in *2009 Conf. Lasers Electro-Optics 2009 Conf. Quantum Electron. Laser Sci. Conf.*, 1–2 (2009).

[16] H. Zhang, C. Li, X. Tu, *et al.*, "Efficient silicon nitride grating coupler with distributed Bragg reflectors," *Opt. Express* 22, 21800–21805 (2014).

[17] C. R. Doerr, L. Chen, Y. K. Chen, *et al.*, "Wide Bandwidth Silicon Nitride Grating Coupler," *IEEE Photonics Technol. Lett.* 22, 1461–1463 (2010).

[18] R. Waldhausl, B. Schnabel, P. Dannberg,¨ *et al.*, "Efficient Coupling into Polymer Waveguides by Gratings," *Appl. Opt.* 36, 9383–9390 (1997).

[19] D. Benedikovic, P. Cheben, J. H. Schmid, *et al.*, "High-efficiency single etch step apodized surface grating coupler using subwavelength structure," *Laser Photon. Rev.* 8(6), L93—-L97 (2014).

[20] Y. Ding, C. Peucheret, H. Ou, *et al.*, "Fully etched apodized grating coupler on the SOI platform with -0.58 dB coupling efficiency," *Opt. Lett.* 39, 5348–5350 (2014).